# Memetic Multi-Objective Particle Swarm Optimization-Based Energy-Aware Virtual Network Embedding

Ashraf A. Shahin[1,2]

[1]College of Computer and Information Sciences,
Al Imam Mohammad Ibn Saud Islamic University (IMSIU)
Riyadh, Kingdom of Saudi Arabia
[2]Department of Computer and Information Sciences, Institute of Statistical Studies & Research,
Cairo University,
Cairo, Egypt
ashraf_shahen@ccis.imamu.edu.sa

*Abstract*—In cloud infrastructure, accommodating multiple virtual networks on a single physical network reduces power consumed by physical resources and minimizes cost of operating cloud data centers. However, mapping multiple virtual network resources to physical network components, called virtual network embedding (VNE), is known to be NP-hard. With considering energy efficiency, the problem becomes more complicated. In this paper, we model energy-aware virtual network embedding, devise metrics for evaluating performance of energy aware virtual network-embedding algorithms, and propose an energy aware virtual network-embedding algorithm based on multi-objective particle swarm optimization augmented with local search to speed up convergence of the proposed algorithm and improve solutions quality. Performance of the proposed algorithm is evaluated and compared with existing algorithms using extensive simulations, which show that the proposed algorithm improves virtual network embedding by increasing revenue and decreasing energy consumption.

*Keywords*—*energy-efficient resource management; green computing; virtual network embedding; cloud computing; resource allocation; substrate network fragmentation*

## I. INTRODUCTION

Cloud computing is a model for enabling on-demand network access to a shared pool of configurable computing resources that can be rapidly provisioned and released with minimal management effort [1]. Cloud computing data centers are established as large-scale data centers containing thousands of servers, switches, and routers that consume enormous amounts of electrical energy and release $CO_2$.

One of the most prominent approaches to address energy inefficiency is to leverage the capabilities of the virtualization technology, which allows creation of multiple Virtual networks on a single physical network [2]. However, mapping virtual resources to physical resources is known to be nondeterministic polynomial-time hard (NP-hard), even if energy efficiency is not considered.

Main objectives of energy-aware virtual network embedding are increasing revenue of substrate network and decreasing power consumed by substrate resources. Revenue can be maximized by increasing number of accommodated virtual networks and decreasing cost of embedding each virtual network. Number of accepted and accommodated virtual networks can be increased by using suitable search technique to find sub-substrate network for accommodating virtual network in reasonable time, regardless of virtual network size or substrate network size. Furthermore, number of accepted virtual networks can be increased by reducing substrate resources fragmentation. Substrate resources are considered fragmented if there are enough substrate resources to achieve virtual network request but virtual network request is rejected due to substrate resources scattering.

Virtual network embedding cost is the total substrate resources used to achieve virtual network request. Virtual network embedding solution maps each virtual node to a substrate node and each virtual link to a loop-free substrate path, which is consists of a set of substrate links. Fig. 1 shows an example of virtual network embedding. The cost of embedding virtual network can be minimized by decreasing number of required substrate links. This can be done by minimizing the length of required substrate paths or by accommodating more than one virtual node from the same virtual network on the same substrate node to eliminate the cost of embedding virtual links between them.

Power consumed by substrate network can be reduced by minimizing number of substrate nodes that are turned *on* from *off* to accommodate virtual node or to participate in substrate path. Furthermore, power can be minimized by selecting substrate nodes that have less power consumption. As shown in Fig. 2, different types of servers have different power consumption rates. Proposing energy-aware virtual network embedding with considering all of the above concerns is a very complicated task.

Multi-objective Particle Swarm Optimization (MOPSO) is a heuristic search technique for optimizing multi-objective optimization problems, which have more than one objective function, such as energy-aware virtual network embedding problem. In such problems, there is no single optimal solution.

Instead, we try to find a set of good solutions that compromise among all objective functions.

In this paper, we propose a model for energy-aware virtual network embedding, devise an energy-aware virtual network embedding metrics to compare different algorithms, and propose memetic multi-objective particle swarm optimization-based energy-aware virtual network embedding algorithm, called MOPSO-EVNE. Performance of the proposed algorithm have been evaluated using extensive simulations, which show that the proposed algorithm increases the long-term average revenue and decreases the power consumption compared with some of existing algorithms.

The remaining of the paper is organized as follows. Section 2 introduces the fundamentals of the proposed algorithm. In Section 3, we discuss the related work on energy-aware virtual network embedding problem. Section 4 presents the virtual network embedding model and problem formulation. Section 5 describes the proposed algorithm. Section 6 evaluates the proposed energy aware virtual network-embedding algorithm using extensive simulations. Finally, in Section 7 we conclude this paper.

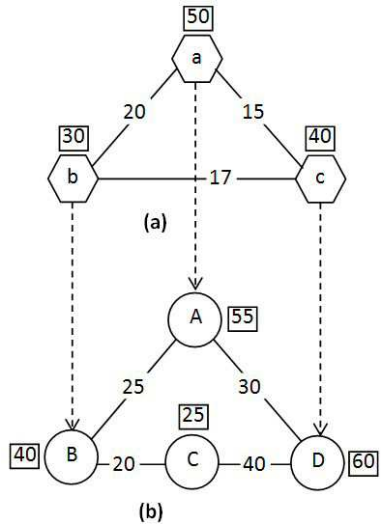

Fig. 1. Virtual network empedding example

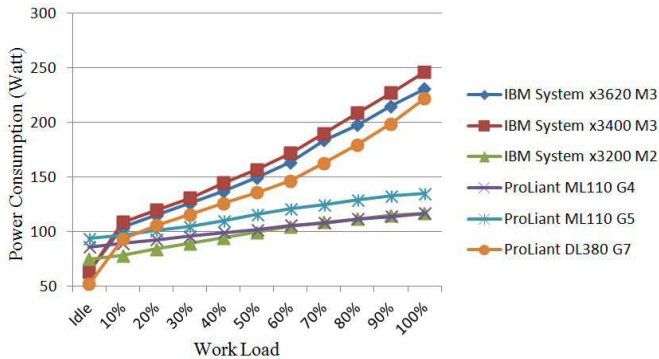

Fig. 2. Power consumption of different types of servers

## II. BACKGROUND

### A. Particle swarm optimization

Particle swarm optimization (PSO) is a population-based stochastic global optimization technique first proposed by Kennedy and Eberhart in 1995 [3]. PSO is inspired by the sociological behavior associated with bird flocking or fish schooling. PSO searches for a possible solution in multiple areas simultaneously and can obtain better optimal solution quickly in lesser computing time than other population based methods. PSO algorithm simultaneously maintains a number of particles, which represent candidate solutions in the search space.

Each particle has position vector and velocity vector, which can be represented as: $X_i(t) = (x_i^1(t), x_i^2(t), \dots, x_i^n(t))$ and $V_i(t) = (v_i^1(t), v_i^2(t), \dots, v_i^n(t))$, where $X_i(t)$ is the position of particle $i$ at time $t$, $V_i(t)$ is the velocity of particle $i$ at time $t$, and $n$ is the dimensions of the solution space. The position and velocity of each particle are updated using the following equations:

$$V_i(t+1) = wV_i(t) + c_1r_1(\text{pBest}_i(t) - X_i(t)) + c_2r_2(\text{gBest}(t) - X_i(t))$$

$$X_i(t+1) = X_i(t) + V_i(t+1)$$

Where $r_1$ and $r_2$ are two random numbers between 0 and 1. The constants $w, c_1,$ and $c_2$ ($0 \le w \le 1.2, 0 \le c_1 \le 2,$ and $0 \le c_2 \le 2$) are specified by user. $\text{pBest}_i(t)$ is the best previous position for the particle i at time t and is known as the personal best position. $\text{gBest}_i(t)$ is the best position among all previous personal best positions at time t and is known as the global best position. The constant w is called *inertia weight* and the first term $wV_i(t)$ is called *inertia component*, which keeps the particle moving forwarding. The constant $c_1$ is called *cognitive weight* and the first term $c_1r_1(\text{pBest}_i(t) - X_i(t))$ is called *cognitive component*, which represents the attraction that a particle has toward its best previous position. The constant $c_2$ is called *social weight* and the first term $c_2r_2(\text{gBest}(t) - X_i(t))$ is called *social component*, which represents the attraction that a particle has toward the global best position. The random numbers $r_1$ and $r_2$ cause the particle to move in a semi-random manner.

### B. Multi-objective optimization

Multi-objective optimization problem can be described as following [4]:

$$\text{Minimize } \vec{f}(\vec{x}) = (f_1(\vec{x}), f_2(\vec{x}), \dots, f_k(\vec{x}))$$

subject to:

$$g_i(\vec{x}) \le 0 \quad i = 1, 2, \dots, m$$

$$h_i(\vec{x}) = 0 \quad i = 1, 2, \dots, p$$

Where $\vec{x} = (x_1, x_2, \dots, x_n)\epsilon \vec{X} \subset \mathbb{R}^n$ is a decision vector consists of $n$ decision variables, $\vec{X}$ is a decision space, $\vec{f}(\vec{x}) = (f_1(\vec{x}), f_2(\vec{x}), \dots, f_k(\vec{x}))\epsilon \vec{Y} \subset \mathbb{R}^k$, is an objective vector consists of $k$ objective functions, $\vec{Y}$ is an objective space, $f_i(\vec{x}): \mathbb{R}^n \to \mathbb{R}, i = 1, 2, \dots, k$ are the objective functions, $g_i(\vec{x}): \mathbb{R}^n \to \mathbb{R}, i = 1, 2, \dots, m$, $h_i(\vec{x}): \mathbb{R}^n \to \mathbb{R}, i = 1, 2, \dots, p$ are inequality and equality constraints functions of the problem. Multi-objective optimization problem tries to find the decision vector $\vec{x}$ in the decision space $\vec{X}$ that will optimize the objective vector $\vec{f}(\vec{x})$.

*Definition 1 (Feasible Solution Set).* The set of all decisions from a decision space $\vec{X}$ that satisfy all inequality and equality constraints is called a feasible solution set, denoted by $\overrightarrow{X_f}$ and $\overrightarrow{X_f} \subset \vec{X}$.

*Definition 2 (Pareto dominance).* Let $\overrightarrow{x^1}, \overrightarrow{x^2} \in \overrightarrow{X_f}$, we say that $\overrightarrow{x^1}$ dominates $\overrightarrow{x^2}$ (denoted by $\overrightarrow{x^1} \prec \overrightarrow{x^2}$) or $\overrightarrow{x^2}$ is dominated by $\overrightarrow{x^1}$ iff $f_i\left(\overrightarrow{x^1}\right) \le f_i\left(\overrightarrow{x^2}\right) \ \forall \ i \in \{1, 2, .., k\} \wedge \exists \ i \in \{1, 2, .., k\} : f_i\left(\overrightarrow{x^1}\right) < f_i\left(\overrightarrow{x^2}\right)$

*Definition 3 (Non-domination).* We say that a decision vector $\vec{x} \in \vec{X}$ is non-dominated with respect to $\vec{X}$, if $\nexists \ \overrightarrow{x'} \in \vec{X}$ such that $\overrightarrow{x'} \prec \vec{x}$.

*Definition 4 (Pareto optimality).* A decision vector $\overrightarrow{x^*} \in \overrightarrow{X_f}$ is Pareto optimal if $\overrightarrow{x^*}$ is non-dominated with respect to $\overrightarrow{X_f}$.

*Definition 5 (Pareto optimal set).* The set of all Pareto optimal decision vectors is called Pareto optimal set and is denoted by $\mathcal{P}^*$.

*Definition 6 (Pareto front).* The Pareto front $\mathcal{PF}^*$ is defined by: $\mathcal{PF}^* = \left\{ \vec{f}(\vec{x}) \ \middle| \ \vec{x} \in \mathcal{P}^* \right\}$

### C. Multi-objective particle swarm optimization

In multi-objective PSO (MOPSO), instead of finding single solution (global best solution), we aim to find a Pareto optimal set, which will be stored in an external repository, called *external archive* [5]. Instead of using global best solution to guide other solutions, Pareto optimal is selected from the *external archive* to guide each particle.

## III. RELATED WORK

In the past few years, several researches have been proposed for effective virtual network embedding and energy-aware virtual network embedding. Rodriguez et al. [6] proposed an integer linear programming model for VNE problem to minimize energy and bandwidth consumption. Rodriguez et al. assigned variant weight values to balance minimization of energy and bandwidth consumption. Their simulation results showed that considering energy consumption minimization only could extremely increase bandwidth consumption and decrease the quality of service; while assigning equal weights to both consumptions minimizes the energy consumption near to optimal solution without significantly increase the bandwidth consumption.

Tarutani et al. [7] studied the energy consumption of the data centers network, which are constructed of optical cross connects and electronic switches (called top-of-rack). Tarutani et al. proposed a virtual network topology called Generalized Flattened Butterfly to achieve sufficient bandwidth and to minimize the energy consumption. The energy consumption is minimized by reducing the number of ports of electronic switches used in the virtual network topology.

Sun et al. [8] modeled the energy-aware VNE problem using mixed-integer programming and proposed a heuristic algorithm to solve the proposed model with efficient power consumption and with minimal violation of service level agreements (SLAs). The proposed algorithm minimizes the energy consumption by consolidating virtual network resources into few substrate resources as possible.

Chang et al. [9] proposed virtual network architecture with virtual network components such as routers and switches. The proposed architecture provides communication functions for virtual resources in Cloud data centers. The authors designed an energy aware routing algorithm for the proposed architecture.

Fischer et al. [10] extended the VNE algorithm proposed by Lischka and Karl in [11] to be energy-aware VNE algorithm. Fischer et al. minimized energy consumption by allowing more than one virtual node from the same virtual network to coexist on the same substrate node. Furthermore, Fischer et al. considered active nodes and nodes that consume less power during node and link mapping to minimize energy consumption.

Beloglazov et al. [12, 13] studied the single VM migration and dynamic VM consolidation problems and they proved the competitive ratios of optimal online deterministic algorithms for energy and performance efficient dynamic VM consolidation. Beloglazov et al. proposed heuristic algorithms for dynamic adaption of VM allocation at run-time based on an analysis of historical data on the resource usage. However, the proposed algorithms do not consider the communication between VMs in allocating or in reallocating VMs.

Cheng et al. [14] proposed topology-aware node ranking technique, called *NodeRank*, to reflect the topological structure of the VNs and the SN. Based on the proposed ranking technique, Cheng et al. proposed two stage virtual network embedding algorithm called *RW-MaxMatch*. However, mapping nodes and links in two independent stages without coordination between them leads to high consumption of the underlying SN's resources. To solve this problem, Cheng et al. [14] proposed RW-BFS algorithm. *RW-BFS* algorithm is a backtracking one-stage VN embedding algorithm, which maps nodes and links at the same stage.

Zhang et al. [15] proposed two VN embedding models: an integer linear programming model and a mixed integer-programming model. Furthermore, Zhang et al. proposed a discrete particle swarm optimization based VNE algorithm, called *RW–PSO*, to solve the proposed models. *RW–PSO* algorithm is an enhanced version of *RW-MaxMatch* [15] algorithm to find near optimal node mapping solutions in large-scale substrate networks. After nodes mapping, Zhang et al. map links using shortest paths algorithm and greedy k-shortest paths algorithm. Cheng et al. [16] proposed discrete Particle Swarm Optimization based virtual network embedding algorithm similar to the proposed algorithm in [15] but they ranked nodes using topology-aware node ranking technique proposed in [14].

Su et al. [17] formulated an energy consumption model for substrate network infrastructures and proposed an extended version of *RW-BFS* [14] algorithm, called *EA-VNE,* for energy-aware virtual network embedding. Su et al. minimized the energy consumption by mapping virtual nodes to Best-fit substrate node according to the required and available CPU to minimize number of active substrate nodes. Virtual links are mapped to shortest loop-free substrate path with minimal number of substrate nodes that are turned on from off.

## IV. VIRTUAL NETWORK EMBEDDING MODEL AND PROBLEM FORMULATION

*Substrate network (SN):* as in our previous work [18, 19], we modeled the substrate network as a weighted undirected graph $G_s = (N_s, L_s)$, where $N_s$ is the set of substrate nodes and $L_s$ is the set of substrate links. Each substrate node $n_s \in N_s$ is weighted by the CPU capacity, and each substrate link $l_s \in L_s$ is weighted by the bandwidth capacity. Fig. 1(b) shows a simple SN example, where the available CPU resources are represented by numbers in rectangles and the available bandwidths are represented by numbers over the links.

*Virtual network (VN):* virtual network $VN_i$ is modeled as a weighted undirected graph $G_{v_i} = (N_{v_i}, L_{v_i})$, where $N_{v_i}$ is the set of virtual nodes and $L_{v_i}$ is the set of virtual links. Virtual nodes and virtual links are weighted by the required CPU and bandwidth, respectively. Fig. 1(a) shows an example of VN with required CPU and bandwidth.

*Virtual network requests (VNR):* the $i^{th}$ VN request $vnr_i$ in the set of all VN requests $VNR$ is modeled as $(G_{v_i}, t_{a_i}, t_{l_i})$, where $G_{v_i}$ is the required VN to be embedded, $t_{a_i}$ is the arrival time, and $t_{l_i}$ is the lifetime. When $vnr_i$ arrives, substrate nodes' CPU and substrate links' bandwidth are allocated to achieve the $vnr_i$. If the substrate network does not have enough resources to achieve $vnr_i$, $vnr_i$ is rejected. At the end of $vnr_i$ lifetime, all allocated resources to virtual $vnr_i$ are released.

*Virtual Network Embedding (VNE):* embedding $VN_i$ on SN is defined as a map $M: G_{v_i} \rightarrow (N'_s, P'_s)$, where $N'_s \subseteq N_s$, and $P'_s \subseteq Path_s$ , where $Path_s$ is the set of all loop free substrate paths in $G_s$. Embedding $VN_i$ can be decomposed into node and link mapping as follows:

Node mapping: $M_N: N_{v_i} \rightarrow N'_s$

Link mapping: $M_L: L_{v_i} \rightarrow P'_s$

*Virtual Network Embedding Revenue:* the revenue of embedding $vnr_i$ at time $t$ is defined as the sum of all required substrate CPU and substrate bandwidth by $vnr_i$ at time $t$.

$$R(vnr_i, t) = Life(vnr_i, t). \left( \sum_{n_{v_i} \in N_{v_i}} CPU(n_{v_i}) + \sum_{l_{v_i} \in L_{v_i}} BW(l_{v_i}) \right) \quad (1)$$

Where $CPU(n_{v_i})$ is the required CPU for the virtual node $n_{v_i}$, $BW(l_{v_i})$ is the required bandwidth for the virtual link $l_{v_i}$, and $Life(vnr_i, t) = 1$ if $vnr_i$ is in its lifetime and substrate resources are allocated to it, otherwise $Life(G_{v_i}, t) = 0$.

*Substrate resources fragmentation (SNF):* substrate resources fragmentation is one of the most important factors that have high impact on VNE revenue and cost. Substrate resources are considered fragmented if there are enough substrate resources to embed VN but the available substrate resources are scattered. VNR will be rejected, because it cannot be allocated to connected substrate resources while there are sufficient substrate resources to achieve this VNR.

Substrate network is considered fragmented if there are two sub-graphs $G_{s_i}, G_{s_j} \subset G_s$, such that $N_{s_i} \cap N_{s_j} = \emptyset$ and $\nexists p'_s \in P'_s$ connects two substrate nodes from $N_{s_i}$ and $N_{s_j}$, where $P'_s$ is the set of all loop free substrate paths in $G_s$ that have available bandwidth greater than or equal a pre-specified lower bound bandwidth and have path length less than or equal a pre-specified maximum path length.

To measure substrate network fragmentation (SNF) at time $t$, we use the following formula:

$$SNF(t) = 1 - \frac{\sum_{i=1}^{m} \left( Residual(G_{s_i}, t) \right)^q}{\left( \sum_{i=1}^{m} Residual(G_{s_i}, t) \right)^q} \quad (2)$$

Where $m$ is the number of fragments in the SN, q is a positive integer number greater than 1 to reduce the influence of the small negligible fragments as long as one large fragment exits, and $Residual(G_{s_i}, t)$ is the total residual substrate resources in sub-substrate network $G_{s_j}$ at time $t$. $Residual(G_{s_i}, t)$ is calculated as following:

$$Residual(G_{s_i}, t) = \sum_{n_{s_i} \in N_{s_i}} CPU_{residual}(n_{s_i}, t) + \sum_{l_{s_i} \in L_{s_i}} BW_{residual}(l_{s_i}, t),$$

$$Where\ G_{s_i} = (N_{s_i}, L_{s_i})$$

The substrate network fragmentation formula in equation (1) is inspired by the fragmentation measure proposed by Gehr and Schneider in [20].

*Virtual Network Embedding Cost:* as in [18, 19], the cost of embedding $vnr_i$ at time $t$ is defined as the sum of all allocated substrate CPU and substrate bandwidth to $vnr_i$ at time $t$.

$$Cost(vnr_i, t) = Life(vnr_i, t). \left( \sum_{n_{v_i} \in N_{v_i}} CPU(n_{v_i}) + \sum_{l_{v_i} \in L_{v_i}} BW(l_{v_i}). Length(M_{L_{v_i}}(l_{v_i})) \right) \quad (3)$$

Where $Length(M_{L_{v_i}}(l_{v_i}))$ is the length of the substrate path that the virtual link $l_{v_i}$ is mapped to.

*Power consumption modeling:* Substrate nodes are turned *on* from *off* to accommodate virtual nodes or to work as intermediate nodes in substrate paths. Recently, there is a new trend to deploy routing cards in data center networks to function as IP routers. Like commercial routers, routing cards handles all packet-processing tasks in hardware with high processing rate and low latency. The power consumption of the routing cards is nearly constant. As shown in [21], fully loading routing card increases its power consumption by around 5% over being idle. As any PCI-based cards, routing card has two states: *enabled* state, which consumes constant power, denoted by $P_r$, and *disabled* state, which does not consume any power.

To model power consumed by substrate nodes to accommodate virtual nodes, we studied the power consumption rates of different types of servers, which are collected using SPEC power benchmark[1] and is depicted in Fig. 2. Fig. 2 shows that each server has a baseline power, which is the power consumed in idle state, and the remaining power consumption is proportional to CPU utilization. Now, we can model the power consumed by an active substrate node $n_s$ at time $t$ as:

$$P_S(n_s, t) = P_b(n_s) + (P_m(n_s) - P_b(n_s)) . CPU(n_s, t) + P_r(n_s) . Routing_{state}(n_s)$$

Where $P_b(n_s)$ is the baseline power of the substrate node $n_s$, $P_m(n_s)$ is the maximum power consumption for the substrate node $n_s$, $CPU(n_s, t)$ is the total CPU utilization for the substrate node $n_s$ at time $t$, $P_r(n_s)$ is the power consumed by active routing card, and $Routing_{state}(n_s)$ is equal to $1$ if the routing card is *enabled* and equal to $0$ if the routing card is *disabled*.

Total power consumed by substrate network at time $t$ is defined as the sum of all power consumed by all substrate nodes at time $t$.

$$Power_S(t) = \sum_{n_s \in N_s} P_S(n_s, t)$$

Power consumption to accommodate virtual node $n_v$ in substrate node $n_s$ can be calculated as following:

$$P_{Vn}(n_v, t) = \begin{cases} P_b(n_s) + (P_m(n_s) - P_b(n_s)) . CPU(n_v), \\ \quad if\ State\ (n_s, t) = 0 \\ (P_m(n_s) - P_b(n_s)) . CPU(n_v), \\ \quad if\ State\ (n_s, t) = 1 \end{cases}$$

Where $State\ (n_s, t)$ is the state of substrate node $n_s$ at time $t$. $State\ (n_s, t)$ equal to $1$ if $n_s$ is *on* and equal to $0$ if $n_s$ is *off*. $CPU(n_v)$ is the required CPU for the virtual node $n_v$.

Power consumption to embed virtual link $l_v$ on substrate path $path_s$ can be calculated as following:

$$P_{Vl}(l_v, t) = \sum_{n_s \in PN_s} \begin{cases} P_b(n_s) + P_r(n_s), \\ \quad if\ State\ (n_s, t) = 0\ and \\ \quad Routing_{state}(n_s) = 0 \\ P_r, \quad if\ State\ (n_s, t) = 1\ and \\ \quad Routing_{state}(n_s) = 0 \\ 0, \quad otherwise \end{cases}$$

Where $PN_s$ is the set of all substrate nodes participate in substrate path $path_s$.

Total power consumption to embed virtual network request $vnr_i$ at time $t$ is defined as the sum of all power consumption to embed its virtual nodes and virtual links.

$$Power_V(vnr_i, t) = \sum_{n_{v_i} \in N_{v_i}} P_{Vn}(n_{v_i}, t) + \sum_{l_{v_i} \in L_{v_i}} P_{Vl}(l_{v_i}, t) \quad (4)$$

*Objectives:* the main objectives are to increase the revenue of VNE, decrease the cost of VNE, decrease the power consumed by substrate nodes, and decrease substrate resources fragmentation in the long run. To evaluate the achievement of these objectives, we use the following metrics:

- *The long-term average revenue*, which is defined by

$$\lim_{T \to \infty} \left( \frac{\sum_{t=0}^{T} \sum_{i=1}^{I} R(vnr_i, t)}{T} \right) \quad (5)$$

Where $I = \| VNR \|$, and $T$ is the total time.

- *The VNR acceptance ratio*, which is defined by

$$\frac{\|VNR_s\|}{\|VNR\|} \quad (6)$$

Where $VNR_s$ is the set of all accepted virtual network requests.

- *The long term R/Cost ratio*, which is defined by

$$\lim_{T \to \infty} \left( \frac{\sum_{t=0}^{T} \sum_{i=1}^{I} R(vnr_i, t)}{\sum_{t=0}^{T} \sum_{i=1}^{I} Cost(vnr_i, t)} \right) \quad (7)$$

- *The long-term average substrate network* fragmentation, which is defined by

$$\lim_{T \to \infty} \left( \frac{\sum_{t=0}^{T} SNF(t)}{T} \right) \quad (8)$$

- *The long-term average substrate network power consumption*, which is defined by

$$\lim_{T \to \infty} \left( \frac{\sum_{t=0}^{T} Power_S(t)}{T} \right) \quad (9)$$

## V. THE PROPOSED ALGORITHM

In this section, we redefine the parameters and operations of the particles in PSO and describe the details of the proposed MOPSO-EVNE algorithm.

### A. Redefining PSO particles operations

We redefined the parameters and operations of the particles in PSO as following:

*Position* (X): the position vector $X_i(t) = (x_i^1(t), x_i^2(t), .., x_i^n(t))$ of a particle $i$ at time $t$ represents virtual node mappings of a VNE solution. $n$ is the number of virtual nodes in the virtual network. All virtual nodes and substrate nodes are ordered and each node has an order number. $x_i^m(t)$ is the order number of substrate node that contains virtual node with order number $m$.

*Velocity* (V): The velocity vector $V_i(t) = (v_i^1(t), v_i^2(t), .., v_i^n(t))$ guides VNE solution (particle) to modifications that enhance current solution. $v_i^m(t)$ is a substrate path specifies a sequence of substrate nodes in which a virtual node with the order number $m$ will be mapped to.

*Subtraction* ($\ominus$): $X_j(t) \ominus X_i(t) = (SPath_{ij}^1(t), SPath_{ij}^2(t), .., SPath_{ij}^n(t))$, where $SPath_{ij}^m(t)$ is a shortest loop free substrate path from substrate node with the order number $x_i^m(t)$ to substrate node with the order number $x_j^m(t)$.

*Addition* ($\oplus$): $p_i V_i(t) \oplus p_j V_j(t)$ indicates that substrate paths are kept from $V_i(t)$ with probability $p_i$ and kept from $V_j(t)$ with probability $p_j$, where $p_i + p_j = 1$.

*Multiplication* ($\otimes$): $X_i(t) \otimes V_i(t+1)$, where $X_i(t) = (x_i^1(t), x_i^2(t), .. \ x_i^m(t).., x_i^n(t))$, and $V_i(t+1) = (v_i^1(t+1), v_i^2(t+1), .., v_i^m(t+1), .., v_i^n(t+1))$ indicates that the virtual node number $m$, which is currently mapped to the substrate node number $x_i^m(t)$, will be mapped to the first substrate node in the substrate path $v_i^m(t+1)$ with enough CPU. If substrate node number $x_i^m(t)$ already participates in the substrate path $v_i^m(t+1)$, the virtual node number $m$ will be mapped to the first substrate node after the substrate node number $x_i^m(t)$ with enough CPU if found.

Finally, position and velocity updating equations are redefined as following:

$$V_i(t+1) = wV_i(t) \oplus c_1 r_1 \left(\text{pBest}_i(t) \ominus X_i(t)\right)$$
$$\oplus c_2 r_2 \left(X_{leader}(t) \ominus X_i(t)\right) \quad (10)$$

$$X_i(t+1) = X_i(t) \otimes V_i(t+1) \quad (11)$$

Where $w + c_1 r_1 + c_2 r_2 = 1$, and $X_{leader}(t)$ is the position vector of the particle (VNE solution) that is used to guide another particle towards better areas in the solution space. According to the redefined operations, $\left(\text{pBest}_i(t) \ominus X_i(t)\right)$ is a set of substrate paths from current position $X_i(t)$ to the personal best position $\text{pBest}_i(t)$, and $\left(X_{leader}(t) \ominus X_i(t)\right)$ is a set of substrate paths from current position $X_i(t)$ to the leader position $X_{leader}(t)$. As a result, $V_i(t+1)$ is a set of substrate paths that guide particle to its personal best position or to position of Pareto optimal solution. The multiplication operation in equation (11) moves each dimension in the position vector $X_i(t)$ one step toward personal best position or toward Pareto optimal solution.

### B. MOPSO-EVNE algorithm

The steps of the proposed multi-objective particle swarm optimization energy aware virtual network-embedding algorithm (MOPSO-EVNE), are shown in Algorithm 1.

Particle swarm $S(t)$ is initialized by collecting a set of VNE feasible solutions. MOPSO-EVNE algorithm initializes $S(t)$ by creating a candidate substrate node list for the virtual node with the largest resources. Candidate substrate nodes list is created by collecting all substrate nodes with enough resources to embed virtual node. Candidate substrate nodes list is sorted in ascending order according to the power consumption rate for each node. Active substrate nodes with lower power consumption are selected first before activating inactive nodes. MOPSO-EVNE visits candidate substrate nodes in the created list sequentially and maps virtual network (starting from the virtual node with the largest resources). Virtual link mappings are performed during the node mapping process in breadth-first search manner to find shortest loop free substrate path with minimum number of activated substrate nodes. MOPSO-EVNE algorithm incrementally increases the maximum allowed substrate path length to visit large number of candidate substrate nodes and maximize the spread of solutions found.

If the *Create_New_Particle*() function failed in creating new VNE feasible solution from the current candidate substrate node, we move to the next candidate node. After initializing particle swarm $S(t)$, each position vector for each particle is improved by using *Improve*(), which applies local search. Each dimension in the particle position vector is remapped to another substrate node, if this mapping improves position vector. New substrate node is specified by creating breadth first search trees from all substrate nodes contains neighbor of the current virtual node. All trees are increased concurrently and the first common substrate node is used as optimization position. Dimensions in the particle position vector are visited in a round robin fashion until no further improves are reached.

In line 29, each particle position vector is evaluated using objective functions specified by equations 1, 2, 3, and 4. Velocity vectors are initialized randomly for each particle. In line 30, $S(t)$ is sorted into a hierarchy of non-dominated Pareto fronts by applying *Fast Nondominated Sorting* approach proposed in [22]. Each particle is assigned a rank value based on its dominance level and crowding distance value.

External archive EA(t) is used to keep the non-dominated solutions found during the search process. External archive solutions will be used as leaders to update velocity vectors of the particles of the swarm. Furthermore, the final output of the MOPSO-EVNE algorithm will be selected from the solutions contained in external archive. In line 32, initial external archive EA(t) is created and the non-dominated solution of the particle swarm S(t) are copied into the external archive EA(t).

Lines from 33 to 47 describe details of each iteration. In each iteration, one of the non-dominated particles is selected from $EA(t)$ to be used as leader. Velocity vector and position vector are updated using equations (10) and (11). To avoid swarm stagnation, position vector is mutated with mutation probability $pro_{mut}$. Without mutation, the proposed algorithm might stop or converge to a local optimum. Mutation is performed by remapping mutated dimension in the position vector to substrate node with enough substrate resources. Virtual links are remapped without considering the maximum substrate path length. *Improve*() function is used to optimize the new position vector to become visible solution. Each particle is evaluated using objective functions and its *pBest* is updated accordingly.

At the end of each iteration, external archive EA(t) must be updated to add new non-dominated solutions found during this iteration. Solutions in external archive EA(t) are combined with the updated solutions in swarm S(t+1), sorted into non-dominated Pareto fronts, and sorted in descending order according to their Crowding-distance values. External archive EA(t+1) is updated by selecting the first $EA_{MaxSize}$ solutions.

After a certain number of iterations, the MOPSO-EVNE algorithm selects best Pareto optimal front from the external archive EA(t) and returns it as suggested solution.

ALGORITHM 1: The details of the MOPSO-VNE algorithm

**INPUTS**:

$G_v = (N_v, L_v)$: VN to be embed

$G_s = (N_s, L_s)$: SN to embed on

$Iterations_{Max}$: maximum number of iterations

$Swarm_{Size}$: swarm size

$EA_{MaxSize}$: maximum size of the external archive

$Max\_backtrack$: upper bound of nodes re-mapping operation

$Hops_{Max}$: maximum allowed substrate path length

**OUTPUTS**:

$M(G_v)$: map VN nodes and links to SN's resources

$S\_VNE$: VN embedding success flag

**Begin**

1: Build breadth-first searching tree of $G_v$ from virtual node with
   largest resources.

2: Sort all nodes in each level in the created breadth-first tree in
   descending order according to their required resources.

3: Create an empty particle swarm $S(t)$ at $t = 0$

4: $Hops = 0$, where $Hops$ is the maximum allowed substrate
   path length in current iteration

5: Build candidate substrate node list $C_s$ for $G_{v_{root}}$

6: **while** size of $S(t) < Swarm_{Size}$ **and** $Hops \leq Hops_{Max}$

7:   **for** each substrate node $n_{s_j} \in C_s$

8:     Create new map $M'(G_v) = \phi$

9:     $Add\left(\left(n_{v_i}, n_{s_j}\right), M'(G_v)\right)$, where $n_{v_i} = G_{v_{root}}$

10:    $backtrack\_count = 0$

11:    **if** $Create\_New\_Particle(G_s, n_{v_{i+1}}, S(t), M'(G_v))$ **then**

12:     $S(t) = S(t) \cup \{M'(G_v)\}$

13:    **else**

14:     $Delete\left(\left(n_{v_i}, n_{s_j}\right), M'(G_v)\right)$

15:    **end if**

16:    **if** size of $S(t) \geq Swarm_{Size}$ **then**

17:     **break**

18:    **end if**

19:   **end for**

20:   $Hops = Hops + 1$

21: **end while**

22: **if** size of $S(t) = 0$ **then**

23:   $S\_VNE = false$

24:   **return**

25: **else**

26:   $Swarm_{Size} = $ size of $S(t)$

27: **end if**

28: $Improve(S(t))$

29: Evaluate each particle in $S(t)$ according to the objective
   functions (1), (2), (3), and (4)

30: Initialize the velocity vector randomly for each particle

31: Sort swarm $S(t)$ into different non-domination levels.

32: create and initialize *external archive* $EA(t)$ *with* non-
   dominated particles in $S(t)$

33: **while** $t < Iterations_{Max}$

34:   **for** each particle $p$ in $S(t)$

35:     Randomly select a single *leader* out of $EA(t)$

36:     Update the particle's velocity vector and the position
       vector using equations (10) and (11).

37:     Perform mutation on particle $p$ with the mutation
       probability $pro_{mut}$

38:     locally improve the particle $p$

39:     Evaluate the particle $p$ according to the objective
       functions (1), (2), (3), and (4)

40:     Update $pBest$ of the particle $p$

41:   **end for**

42:   Sort all particles in $S(t + 1) \cup EA(t)$ into different non-
     domination levels.

43:   Calculate Crowding-distance for each particle in $S(t + 1) \cup$
     $EA(t)$

44:   Sort in $S(t + 1) \cup EA(t)$ *in descending order based on*
     Crowding-distance values

45:   Update external archive $EA(t + 1)$ by getting the first
     $EA_{MaxSize}$ particles from the sorted $S(t + 1) \cup EA(t)$

46:   $t = t + 1$

47: **end while**

48: $M(G_v) = Best\_Pareto\_optimal\_front(EA(t))$

49: $S\_VNE = true$

50: **return**

**End**

## VI. PERFORMANCE EVALUATION

To evaluate the performance of the proposed algorithm, we have compared its performance with the following algorithms: *RW-MaxMatch* [16], *RW-BFS* [14], *AdvSubgraph-MM* [10], *AdvSubgraph-MM-EE* [10], and *AdvSubgraph-MM-EE-Link* [10]. In the following subsections, we describe the evaluation environment settings and discuss the simulations' results.

### A. Evaluation environment settings

Performance is evaluated using two substrate network topologies, which are generated using Waxman generator. The first SN topology is configured with 50 nodes and 250 links. Bandwidth of the substrate links are uniformly distributed between 50 and 100 with average 75. The second SN topology is configured with 200 nodes and 1000 links. Bandwidth of the substrate links are uniformly distributed between 50 and 150 with average 100. Each substrate node is randomly assigned one of the following server configurations: HP ProLiant ML110 G4 (Intel Xeon 3040, 2 cores X 1860 MHz, 4 GB), or HP ProLiant ML110 G5 (Intel Xeon 3075, 2 cores X 2660 MHz, 4 GB).

We generated 1000 Virtual network topologies using Waxman generator with average connectivity 50%. The number of virtual nodes in each VN is variant from 2 to 20. Each virtual node is randomly assigned one of the following CPU: 2500 MIPS, 2000 MIPS, 1000 MIPS, and 500 MIPS, which are correspond to the CPU of Amazon EC2 instance types. Bandwidths of the virtual links are real numbers uniformly distributed between 1 and 50. VN's arrival times are generated randomly with arrival rate 10 VNs per 100 time units. The lifetimes of the VNRs are generated randomly between 300 and 700 time units with average 500 time units. Generated SN and VNs topologies are stored in brite format and used as inputs for all algorithms. For all algorithms, we set the maximum allowed hops ($Hops_{Max}$) to 2, and the upper bound of remapping process ($Max\_backtrack$) to $3n$, where $n$ is the number of nodes in each VNR. $Iterations_{Max}$ and Swarm$_{Size}$ of the MOPSO-EVNE algorithm are set to 5 and 10. Finally, we compared the results from the implemented algorithms.

## B. Evaluation results

MOPSO-EVNE algorithm increases VNR acceptance ratio as shown in Fig. 3 and Fig. 4. Fig. 3 shows the VNR acceptance ratio comparison using the first substrate network, which is configured with 50 substrate nodes and 250 virtual links. Fig. 4 shows the VNR acceptance ratio comparison using the second substrate network, which is configured with 200 substrate nodes and 1000 virtual links. *AdvSubgraph-MM*, *AdvSubgraph-MM-EE*, and *AdvSubgraph-MM-EE-Link* are not compared using the second substrate network (200 nodes) because they have high complexity (require more than one month).

VNR acceptance ratio is evaluated using equation (6), which only considers the number of accepted VNRs without considering variations between VNRs' sizes. In Fig. 5 and Fig. 6, we compared the ratio of accepted virtual resources (virtual CPU and virtual BW) without considering its VNRs.

Although, MOPSO-EVNE algorithm increases the acceptance ratio among other algorithms, it rejects 81% and 33% of virtual resources (Fig. 7 and Fig. 8). The reason behind this rejection is the lack of available substrate resources (Fig. 9 and Fig. 10), especially the lack of available substrate CPU (Fig. 11 and Fig. 12).

MOPSO-EVNE algorithm increases the long-term average revenue, which is defined by equation (5) (Fig. 13 and Fig. 14). As shown in Fig. 15 and Fig. 16, MOPSO-EVNE algorithm increases the revenue compared with the cost of embedding VNRs. In Fig. 15, revenue/cost ratio of MOPSO-EVNE algorithm exceeds 100%, which means that the cost of embedding VNRs is less than gained revenue from embedding them. MOPSO-EVNE algorithm increases the revenue by increasing substrate resource utilization (Fig. 17 and Fig. 18) and reducing substrate resources fragmentation (Fig. 19), which is defined by equation 8.

The long-term average substrate network power consumption is compared and depicted in Fig. 20 and Fig. 21. Fig. 20 and Fig. 21 show that MOPSO-EVNE algorithm consumes more power, but this is due to the large amount of accommodated virtual resources. To investigate this point, we compared the power consumed by accommodating one unit of virtual resources. Fig. 22 and Fig. 23 show the comparison results. *RW-MaxMatch* algorithm is removed from Fig. 22 because it has a very high power consumption rate. Although, MOPSO-EVNE algorithm activated more substrate nodes to achieve more VNRs (Fig. 24 and Fig. 25), the power consumption rate of the proposed algorithm is similar to the power consumption rate of the *AdvSubgraph-MM-EE-Link* algorithm using small substrate network. However, *AdvSubgraph-MM-EE-Link* algorithm is not applicable to large substrate networks.

Although, we run our simulation with small size of particle swarm (10 particles) and with small number of iterations (5 iterations), MOPSO-EVNE algorithm increases the revenue and the acceptance ratio in reasonable time. Fig. 26 and Fig. 27 show the average VNE time consumed by each algorithm.

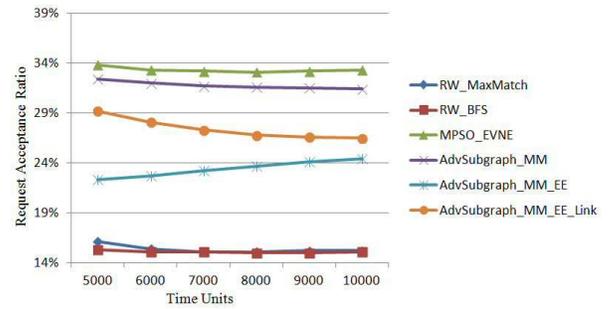

Fig. 3. VNR acceptance ratio comparison using 50 substrate nodes

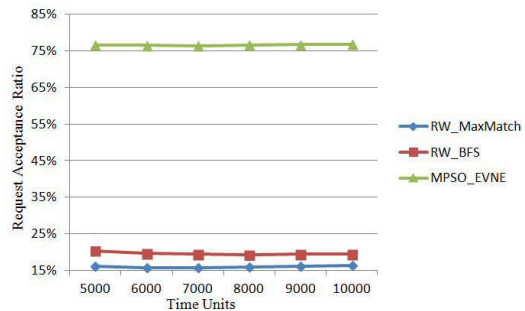

Fig. 4. VNR acceptance ratio comparison using 200 substrate nodes

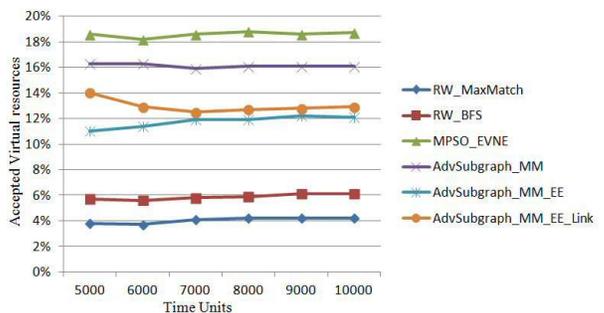

Fig. 5. Virtual resources acceptance ratio comparison using 50 substrate nodes

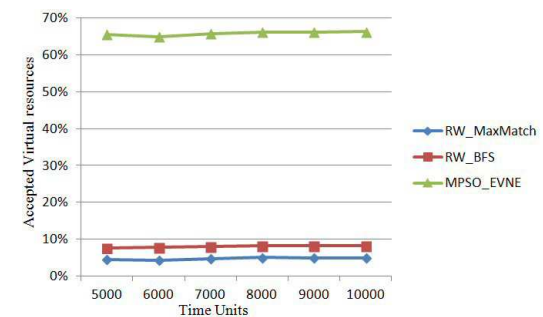

Fig. 6. Virtual resources acceptance ratio comparison using 200 substrate nodes

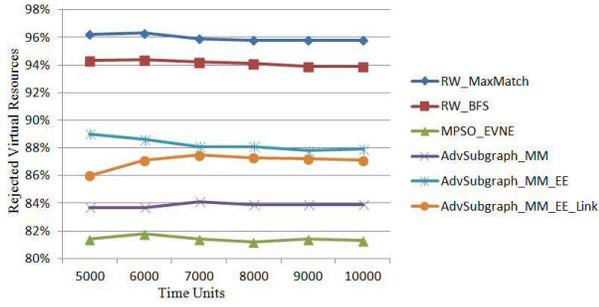

Fig. 7. Rejected virtual resources comparison using 50 substrate nodes

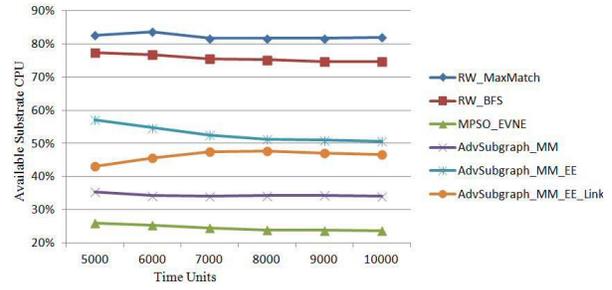

Fig. 11. Available substrate CPU comparison using 50 substrate nodes

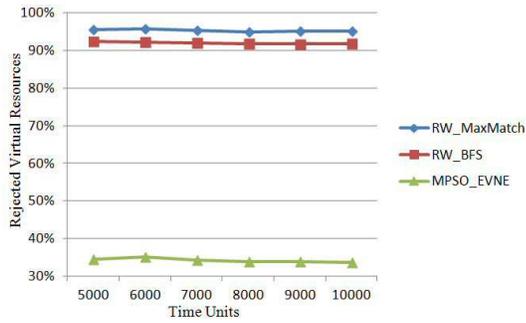

Fig. 8. Rejected virtual resources comparison using 200 substrate nodes

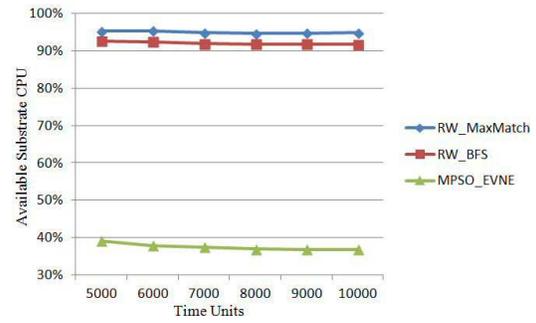

Fig. 12. Available substrate CPU comparison using 200 substrate nodes

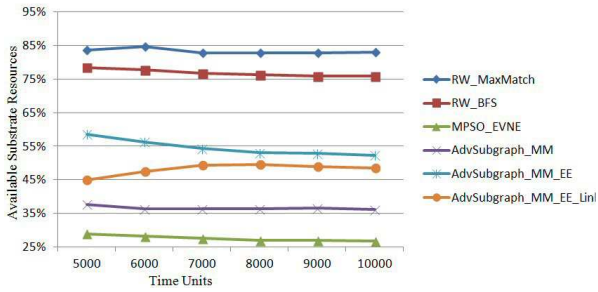

Fig. 9. Available substrate resources comparison using 50 substrate nodes

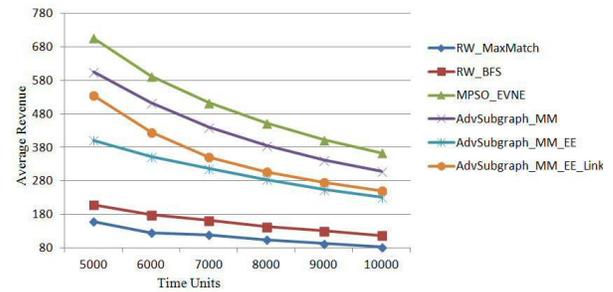

Fig. 13. Revenue comparison using 50 substrate nodes

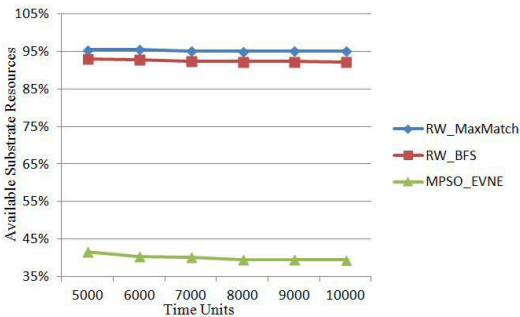

Fig. 10. Available substrate resources comparison using 200 substrate nodes

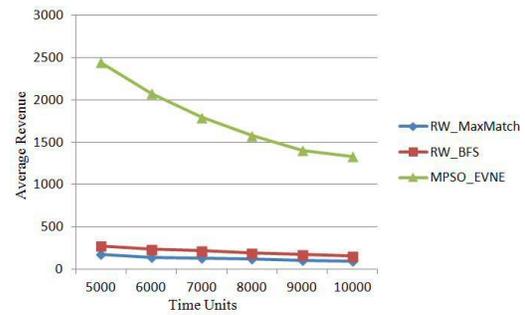

Fig. 14. Revenue comparison using 200 substrate nodes

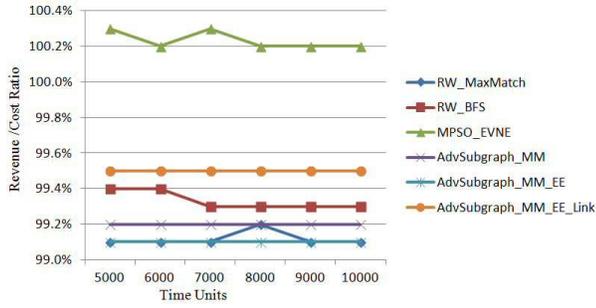

Fig. 15. Revenue/Cost ratio comparison using 50 substrate nodes

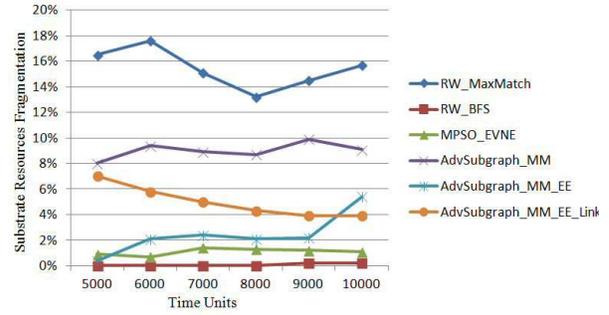

Fig. 19. Substrate resources fragmentation comparison using 50 substrate nodes

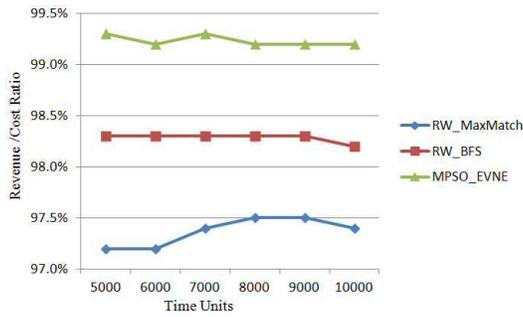

Fig. 16. Revenue/Cost ratio comparison using 200 substrate nodes

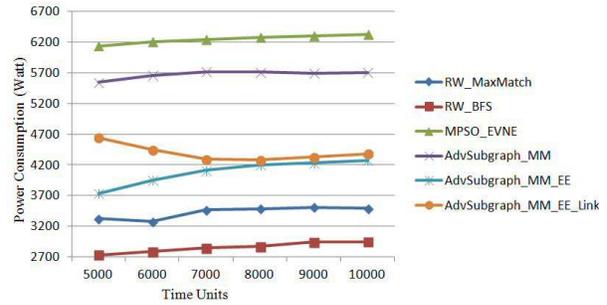

Fig. 20. Power consumption comparison using 50 substrate nodes

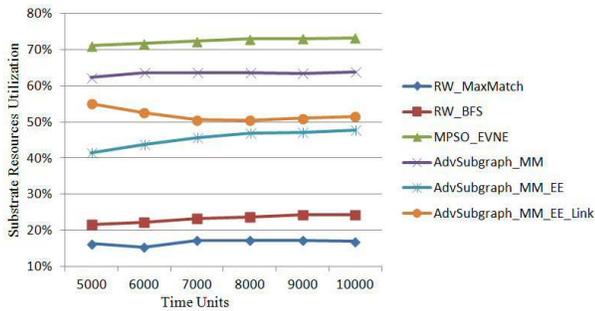

Fig. 17. Substrate resources utilization comparison using 50 substrate nodes

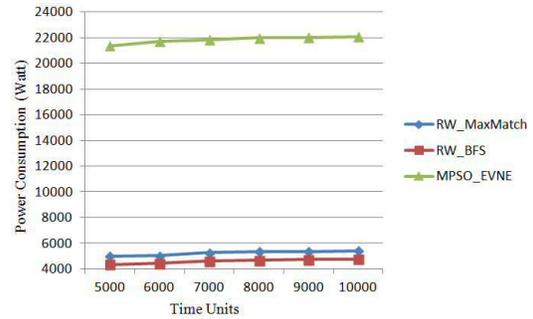

Fig. 21. Power consumption comparison using 200 substrate nodes

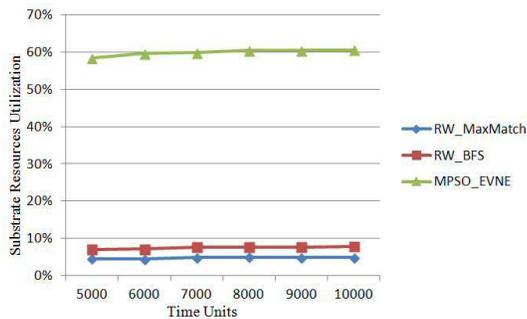

Fig. 18. Substrate resources utilization comparison using 200 substrate nodes

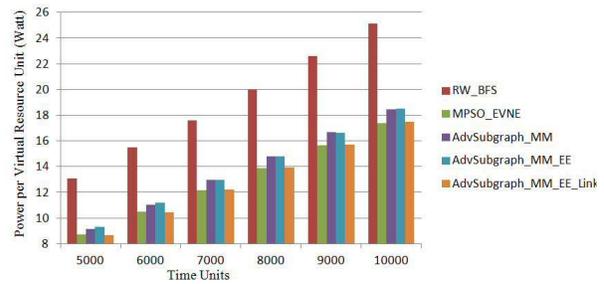

Fig. 22. Comparing power consumption per virtual resource unit using 50 substrate nodes

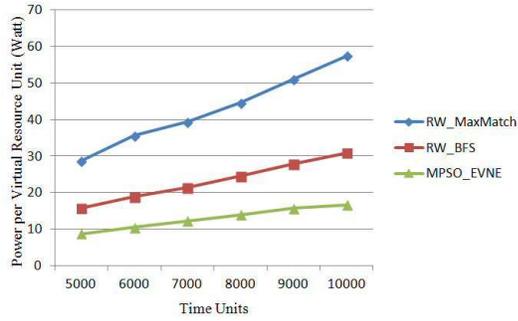

Fig. 23. Comparing power consumption per virtual resource unit using 200 substrate nodes

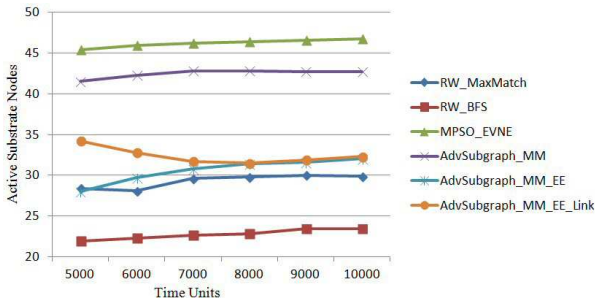

Fig. 24. Active substrate nodes comparison using 50 substrate nodes

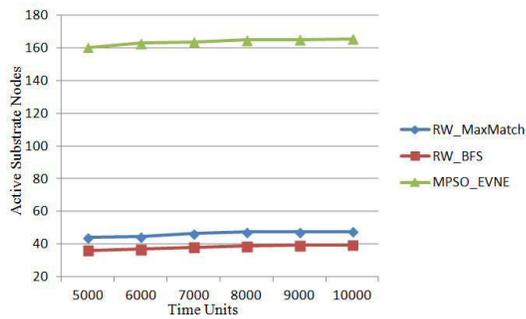

Fig. 25. Active substrate nodes comparison using 200 substrate nodes

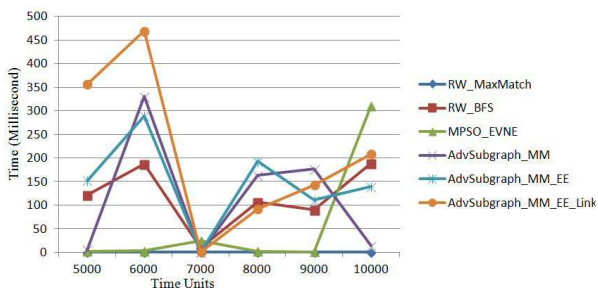

Fig. 26. Average virtual network embedding time comparison using 50 substrate nodes

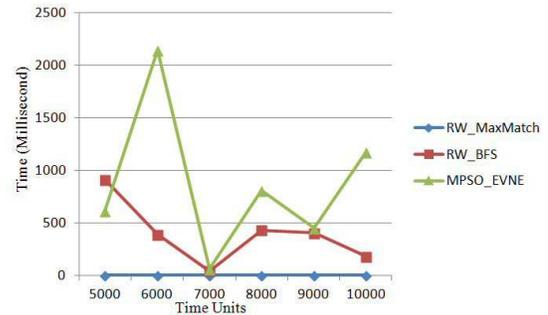

Fig. 27. Average virtual network embedding time comparison using 200 substrate nodes

## VII. CONCLUTION

Embedding multiple virtual networks on a shared substrate network is NP-hard. This complexity is increased by considering energy efficiency of virtual network embedding. In this paper, we modeled energy-aware virtual network embedding problem and proposed an efficient energy aware virtual network-embedding algorithm based on multi-objective particle swarm optimization. The proposed algorithm aims to find good "tradeoff" virtual network embedding solutions that represent the best possible compromises among virtual network embedding revenue, cost, fragmentation, acceptance, and power consumption. Local search is employed to enhance position vector of each particle and to speed up the convergence of the proposed algorithm. Elitism is insured by storing best non-dominated virtual network embedding solutions into external archive. Extensive simulations show that the proposed algorithm outperforms previous algorithms in terms of the long-term average revenue, long-term average cost, long-term average substrate resources fragmentation, and long-term average power consumption. For the future work, we plan to extend the proposed algorithm to consider variant workload and employ virtual machine migration and virtual link migration to enhance energy efficiency of the proposed algorithm.